\documentclass{cplslarge}%
\usepackage[numbers]{natbib}

\begin{document}
\mainmatter
\author[Bouchet \&\ Venaille]{f. bouchet, a. venaille}

\chapter{Zonal flows as statistical equilibria}

The basic question of the theoretical part of this book is "Why do zonal jets form so often in turbulent flows dominated by geostrophic balance, and do they have universal properties ?". Indeed, zonal jets are striking and beautiful examples of the propensity for geophysical turbulent flows to spontaneously self-organize into robust, large scale coherent structures. Strong mid-basin eastwards jets such as the Gulf-Stream or the Kuroshio are observed in any of the Earth ocean in the presence of western boundaries. Several chapters of this book propose dynamical mechanisms for the formation of zonal jets: statistical theories (kinetic approaches, second order or larger oder closures), deterministic approaches (modulational instability, $\beta$-plumes, radiating instability, zonostrophic turbulence, and so on). A striking remark is that all these different dynamical approaches, each of them possibly relevant in some specific regimes, lead to the same kind of final jet structures. Is it  then possible to have a more general explanation of why all these different dynamical regimes, from fully turbulent flows to gentle quasilinear regime, consistently lead to the same jet attractors ?  

Equilibrium statistical mechanics provides an answer to this general question. Starting from first principle, one can build microcanonical measures which are natural invariant measures of the inertial dynamics. Those probability measures can also be interpreted as corresponding averages over set of vorticity field, chosen randomly with fixed energy and other invariants, and with a uniform probability. From a physical point of view, the most important result  is that for a wide range of external parameters such as the energy or other invariants, most of these random vorticity fields have a large scale structure which is very close to a well identified zonal jets. Zonal jets are then understood as the most probable macroscopic organization of the flow.  Moreover as discussed below, many conclusions that generalize to most of geometries and models can be deduced from this equilibrium statistical mechanic approach. This leads to far reaching conclusions on the natural propensity of these turbulent flow to self-organize and onto the way they should do that.

Beside these very positive aspects, the equilibrium statistical mechanics approach has also his own drawbacks.  The main one is that statistical equilibria are parameterized by all conserved quantities. Because of conservation of the potential vorticity, inertial models of geostrophic turbulence usually have a huge amount of such invariants. As a consequence, there is a huge number of statistical equilibria. In situations of weak forces and dissipation (when the spin up or spin down time is much longer than a time scales related to the inertial dynamics), which is very common in geophysics, the actual attractor is selected among all statistical equilibria by the long term balance between forces and dissipations. In such cases, equilibrium statistical mechanics can predict the general structure, but a fully predictive theory must rely on a dynamical approach like ones discussed in other chapters of this book. 

From these preliminary remarks, the aim of this contribution will be, first to present briefly the theoretical ideas related to equilibrium statistical mechanics. Then we will describe the universal conclusions that can be drawn from the properties of statistical equilibria, for the class of quasi-geostrophic models.  We also describe statistical mechanics modeling of a few real geophysical flow and their comparison with observation data. Those model sometimes led to a deeper understanding of the large scale flow structure than the previous approaches. 

We can readily summarize the general conclusions that can be drawn from equilibrium statistical mechanics. First, as soon as the beta-effect or the variation of the Coriolis parameter are taken into consideration, low energy statistical equilibria are generally zonal jets. However, when the energy is increased, or the angular momentum is considered for spherical geometries, we observe transitions to non zonal attractors. This is actually often observed, as for Jupiter where in the south hemisphere the Great Red Spot and other vortices are embedded in the zonal jets. We can actually predict such a transition, with the overall energy as a control parameter. Second, statistical equilibria lead to potential vorticity homogenization consistently with the classical picture 
developed independently of statistical mechanics. Statistical mechanics allow to go further: by characterizing the statistical equilibria as the one with the PV homogenized as much as possible given the energy constraint, it actually provide a quantitative and predictive theory of potential vorticity homogenization. It explains why this homogenization is not complete, and how it should be incomplete, and also explain why we often observe homogenization in large subdomains of the flow but not over the whole domain.  The third general result is the tendency of the flow to barotropization, which is actually a well observed phenomena. But again barotropization should not be expected to be complete, as a complete barotropization is prevented by invariants, the energy and the different distribution of potential vorticity in different layers. A last striking inference of equilibrium statistical mechanics, is that the final state of these complex dynamical phenomena is expected to be described by only a few control parameters.

From an historical point of view, the first attempt to use equilibrium statistical mechanics ideas
to explain the self-organization of 2D turbulence was performed by Onsager 
\citep{onsager1949} in the framework of
the point vortex model. In order to treat flows with continuous vorticity fields, another approach has been proposed by Kraichnan in the framework of the truncated Euler equations \citep{kraichnan1975}, which has then been applied to  quasi-geostrophic flows over topography \citep{salmonhollowayetal1976,carnevalefrederiksen1987}. The truncation has a drastic consequence: only the energy and the enstrophy  are conserved quantities, while any function of the  vorticity is conserved for the Euler equation. 
 There exists now a theory for the Euler or the quasi-geostrophic dynamics, the Miller--Robert--Sommeria
(MRS hereafter) equilibrium statistical mechanics, that explains the
spontaneous organization of unforced and undissipated two-dimensional
and geophysical flows \citep{robert1990,miller1990,robert1991,robertsommeria1991}.
From the knowledge of the energy and the global distribution of potential
vorticity levels provided by an initial condition, this theory predicts
the large scale flow as the most probable outcome of turbulent mixing.

There already exist several presentations of the equilibrium statistical
mechanics of two-dimensional and geostrophic turbulent flows \citep{sommeria2001,limnebus2007,majdawang2006},
some emphasizing kinetic approaches of the point-vortex model \citep{chavanis2002},
other focusing on the legacy of Onsager \citep{eyinksreenivasan2006}
or the applications of statistical mechanics to climate problems \citep{marston2011,lucarinibenderetal2013}.
A precise explanation of the statistical mechanics basis of the MRS
theory, actual computations of a large class of equilibrium macrostates
and further references can be found in \citep{bouchetvenaille2012}. 

We present in the first section statistical mechanics theory of the one
layer quasi-geostrophic models. We summarize in the second section
existing result on the emergence of zonal flows as equilibrium
macrostates on a beta plane when the relation between potential vorticity
and streamfunction is assumed to be linear, and show that breaking
of the zonal symmetry usually appears when increasing the energy of
the flow. We discuss in a third section the case of small Rossby radius
of deformation, which allows to describe equilibrium macrostates presenting
strong mid basin zonal jets such as the Gulf Stream, but also non-zonal
structures such as the Jupiter's Great Red Spot. We explain that a few
parameters such as the energy, the aspect ratio and the vorticity
asymmetry allow to predict which of these macrostates is the equilibrium one. The equilibrium
statistical mechanics has a limited range of applicability since it
neglects the effect of forcing and dissipation, and assume that there
is sufficient mixing in phase space. The validity of these hypothesis
will be discussed in more details in the fourth section.

\section{Statistical mechanics of quasi-geostrophic flows}

\subsection{The quasi-geostrophic models \label{sub:quasi-geostrophic Model}}

We introduce here the quasi-geostrophic barotropic and equivalent barotropic (1-1/2 layer) models. There
are several books discussing these models in more details, among which
\citep{gill1982,pedlosky1982,salmon1998,vallis2006}. The dynamics
of the equivalent barotropic model is the transport of the potential vorticity $q$
by a non-divergent velocity field: 
\begin{equation}
\frac{\partial q}{\partial t}+\mathbf{v}\cdot\nabla q=0,\quad\mathrm{with}\,\,\,\mathbf{v}=\mathbf{e}_{z}\times\nabla\psi\ ,\label{QG}
\end{equation}
\begin{equation}
q=\Delta\psi-\frac{\psi}{R^{2}}+\eta_{d}\ ,\label{dir}
\end{equation}
 where $\eta_{d}(x,y)$ is an equivalent topography, which does not
vary with time, and $R$ is the Rossby radius of deformation, a characteristic
length of the system, and $\psi(x,y,t)$ the streamfunction.

When $R \rightarrow + \infty$, the system of equations  (\ref{QG}-\ref{dir})
is the barotropic model that describes the dynamics of an homogeneous
layer of fluid of depth $H$. In that case, the term $\eta_{d}=\beta y+(f/H)h_{b}$
represents the combined effects of the planetary vorticity gradient
$\beta y$ and of a real bottom topography $h_{b}$. We also
consider in this chapter the case of barotropic flows on a sphere, for which $ $$\eta_{d}=2\Omega\sin\theta+(f/H)h_{b}$,
where $\theta$ is the latitude.

When $R$ is of order one or smaller, the system of equations (\ref{QG}-\ref{dir})
is the equivalent barotropic model, also called "Charney-Hasegawa-Mima"
model or 1-1/2 layer model, which describes the dynamics of an active
layer above a lower denser layer either at rest or characterized by
a prescribed stationary current $\psi_{d}$. In that case, the Rossby
radius is related to the relative density difference $\Delta\rho/\rho$
between both layers, on the gravity $g$, on the Coriolis parameter
$f_{0}$, and on the mean depth $H$ of the upper layer though $R=\left(Hg\Delta\rho/\rho\right)^{1/2}/f_{0}$,
and the term $\eta_{d}=\beta y+\psi_{d}/R^{2}$ represents the combined
effects of the planetary vorticity gradient $\beta y$ and of a the
streamfunction $\psi_{d}$ of the deep layer. 

For the boundary conditions, two cases will be distinguished, depending
on the domain geometry $\mathcal{D}$. In the case of a closed domain,
there is an impermeability constraint (no flow across the boundary),
which amounts to a constant streamfunction along the boundary. To
simplify the presentation, the condition $\psi=0$ at boundaries will
be considered%
\footnote{The physically relevant boundary condition should be $\psi=\psi_{fr}$
where $\psi_{fr}$ is determined by using the mass conservation constraint
$\int\mathrm{d}\mathbf{r}\ \psi=0$ ($\psi$ is proportional to interface
variations). Taking $\psi=0$ does not change quantitatively the solutions
in the domain bulk, but only the strength of boundary jets.%
}. In the case of a zonal channel, the streamfunction $\psi$ is periodic
in the $x$ direction, and the impermeability constraint applies on
northern and southern boundaries. In the remaining two sections, length
scales are nondimensionalized such that the domain area $|\mathcal{D}|$
is equal to one.

\subsection{Dynamical invariants and their consequences \label{sec:2D-Geostrophic-Turbulence}}

According to Noether's Theorem, each symmetry of the system is associated
with the existence of a dynamical invariant, see e.g. \citep{salmon1998}.
These invariants are crucial quantities, because they provide strong
constraints for the flow evolution. Starting from Eq. (\ref{QG}), (\ref{dir})
and the aforementioned boundary conditions one can prove that the quasi-geostrophic
equations conserve the energy: 
\begin{equation}
E=\frac{1}{2}\int_{\mathcal{D}}\mathrm{d}\mathbf{r}\,\left[(\nabla\psi)^{2}+\frac{\psi^{2}}{R^{2}}\right]=-\frac{1}{2}\int_{\mathcal{D}}\mathrm{d}\mathbf{r}\,\left(q-\eta_{d}\right)\psi,\label{ene}
\end{equation}
with $\mathbf{r}=(x,y)$. Additionally, the quasi-geostrophic dynamics
(\ref{QG}) is a transport by an incompressible flow, so that the
area $\gamma\left(\sigma\right)\mathrm{d}\sigma$ occupied by set of points with  $\sigma \leq \omega (\mathbf{r}) \leq \sigma + \mathrm{d}\sigma$, is a dynamical invariant. The quantity $\gamma(\sigma)$
will be referred to as the global distribution of potential vorticity.
The conservation of the distribution $\gamma\left(\sigma\right)$
is equivalent to the conservation of the Casimir's functionals $\int_{\mathcal{D}}\mathrm{d}\mathbf{r}\, f(q)$,
where $f$ is any sufficiently smooth function (see for instance  \citep{bouchetpottersetal2013}). Two important Casimirs
are the potential enstrophy $Z=\int_{\mathcal{D}}\mathrm{d}\mathbf{r}\, q^{2}/2$
and the circulation $\Gamma=\int_{\mathcal{D}}\mathrm{d}\mathbf{r}\, q$.

If the domain where the flow takes place has symmetries, each of those imply the existence of an additional
invariant (for instance angular momentum for the sphere), that should be taken into account.

Fj{\o}rtoft first showed that a striking consequence of the Casimir conservation laws, especially enstrophy, is the fact
that energy has to remain at large scale in freely evolving quasi-geostrophic
dynamics.  This contrasts
with three dimensional turbulence, where the direct energy cascade
(toward small scale) would rapidly dissipate such structures. The formation and  persistence of long lived coherent structures containing most of the energy is actually observed, in experiments, geophysical flows and numerical simulations.  These
long lived coherent structures are often observed to be very close to steady solutions of the quasi-geostrophic
equations (\ref{QG}), that act as attractors. They satisfy $\mathbf{v}\cdot\nabla q=\nabla\psi\times\nabla q=0$.
It means that steady solutions are flows for which streamlines
are isolines of the potential vorticity. For instance, any flow characterized
by a $q-\psi$ functional relationship is a steady solution of the quasi-geostrophic dynamics. 
One can understand the existence of an infinite number of steady solutions as a consequence of the infinite number of invariants (more details and more consequences of the conservations laws in two-dimensional
and geophysical turbulent flows can be found in \citep{bouchetvenaille2012}).
Zonal flows are actually special examples of such steady solution.

At this point, we need a theory i) to support the idea that the freely
evolving flow dynamics will effectively self-organize into a state which is close to a steady solution ii) to determine the $q-\psi$ relationship associated
with this steady solution iii) to select the dynamical equilibria
that are likely to be observed. This is the goal and the achievement
of equilibrium statistical mechanics theory, presented in the next
subsection.

\subsection{The equilibrium statistical mechanics of Miller--Robert--Sommeria (MRS)
\label{sec:Equilibrium-statistical-mechanics}}

The MRS equilibrium statistical theory initially developed by \citep{robert1990,miller1990,robert1991,robertsommeria1991}
is introduced on a heuristic level in the following. There exist rigorous
justifications of the theory, see for instance \citep{bouchetvenaille2012} for
detailed discussions and further references.

The main idea of statistical mechanics is to make the distinction between microscopic states that determine completely the state of the system, and macroscopic states that are sets of microscopic states with the same macroscopic properties. For the quasi-geostrophic models, the potential vorticity field are natural microscopic states as they completely determine the state of the flow. We define a macroscopic state as a local probability probability $\rho(\sigma,\mathbf{r})\mathrm{d}\sigma$ to observe a potential vorticity level $\sigma$ at a point $\mathbf{r}=\left(x,y\right)$.  Such a local probability can be computed for any potential vorticity field $q$ through a coarse-graining procedure, such that many microstates correspond to a unique macrostate. 
The microcanonical measure is defined as the measure where all microscopic states with a given value of the dynamical invariants are equiprobable. The specific Boltzmann entropy of a macrostate is then defined as the logarithm of the number of microstates that have a prescribed value of the dynamical invariants and that correspond to the same macrostate, divided by the number of degrees of freedom (see for instance  \citep{bouchetpottersetal2013} for simple definitions and illustrations in the case of discretized two-dimensional Euler models).

We note that, because $\rho$ is a local probability, and because all microstates should be consistent with the fixed potential vorticity distribution, the relevant macrostates must verify the two constraints:
\begin{itemize}
\item The local normalization $N\left[\rho\right](\mathbf{r})\equiv\int_{\Sigma}\mathrm{d}\sigma\,\,\rho\left(\sigma,\mathbf{r}\right)=1$, 
\item The global potential vorticity distribution $D_{\sigma}\left[\rho\right]\equiv\int_{\mathcal{D}}d\mathbf{r}\,\rho\left(\sigma,{\mathbf{r}}\right)=\gamma\left(\sigma\right)$. 
\end{itemize}
For any macrostate $\rho$, we define the averaged potential vorticity field, by $\overline{q}\left(\mathbf{r}\right)=\int_{\Sigma}\mathrm{d}\sigma\,\,\sigma\rho\left(\sigma,\mathbf{r}\right)$, which is related to the macrostate streamfunction through $\overline{q}=\Delta{\psi}-{\psi}/R^{2}+\eta_{d}$. We can define the macrostate energy as the energy of its averaged  potential vorticity field: $\mathcal{E}\left[\rho\right]\equiv-\frac{1}{2}\int_{\mathcal{D}}\mathrm{d}\mathbf{r}\int_{\Sigma}\mathrm{d}\sigma\,\rho\left({\sigma}-\eta_{d}\right){\psi}=E$. 
At this stage, the careful readers would have noticed that the energies of all microstates of a macrostate $\rho$ may not be equal between each other, and not be equal to the macrostate energy which is the energy of the averaged vorticity field. Identifying such quantities, which are actually different quantities, is classically referred as a mean-field hypothesis in statistical mechanics. However, because an overwhelming number of microstates have only only small scale fluctuations around the mean field potential vorticity, and because energy is a large scale quantity, for most microstates the contributions of these fluctuations to the total energy are negligible with respect to the mean-field energy. Then the mean-field hypothesis is actually exact for the microcanonical ensembles of the 2D-Euler and quasi-geostrophic dynamics. The above heuristic explanation, and the fact that the mean-field approximation is valid, can indeed be made fully rigorous using large deviation theory \citep{michelrobert1994,bouchetellisetal2000} (see an heuristic discussion in \citep{bouchetpottersetal2013}).

The aim of statistical mechanics, and the achievement of the MRS statistical theory for the two-dimensional Euler or quasi-geostrophic models, is the computation of the entropy of the macrostates $\rho$. The result is that the entropy of a macrostate $rho$ is equal to the entropy functional 
\begin{equation}
\mathcal{S}\left[\rho\right]\equiv-\int_{\mathcal{D}}\mathrm{d}\mathbf{r}\int_{\Sigma}\mathrm{d}\sigma\,\rho\log\rho\ ,\label{eq:Entropie_Maxwell_Boltzmann}
\end{equation}
if $\rho$ satisfies the constraints associated with each dynamical invariant (the normalization, the potential vorticity and the mean-field energy constraint), or equal to $-\infty$ otherwise.
By construction, the entropy (\ref{eq:Entropie_Maxwell_Boltzmann}) is a quantification
of the number of microscopic states $q$ corresponding to a given
macroscopic state $\rho$. Then the most probable macrostate, also called equilibrium macrostate, $\rho_{eq}$, is the maximizer of (\ref{eq:Entropie_Maxwell_Boltzmann}) with the constraints corresponding to the dynamical invariants. A crucial remark is that an overwhelming number of microstates actually correspond to the equilibrium macrostate, such that in the limit of an infinite number of degrees of freedom the probability to observe the equilibrium macrostate in the microcanonical measure is actually one \citep{michelrobert1994}. This gives the meaning of the equilibrium macrostate, and explains why the equilibrium macrostate is a natural candidate for the prediction of the final outcome of any turbulent dynamics. While the statement  "for any microcanonical ensemble, an overwhelming number of microstates are part of the same equilibrium macrostate"  is actually a theorem, the statement about the prediction of the final outcome of the evolution of a turbulent dynamics requires further that ergodicity is not broken by the dynamics.

The first step toward the  computation of MRS equilibria is to find critical
points $\rho$ of the mixing entropy (\ref{eq:Entropie_Maxwell_Boltzmann}) with constraints.
In order to take into account the constraints, one needs to introduce
the Lagrange multipliers $\zeta(\mathbf{r})$, $\alpha(\sigma)$,
and $\lambda$ associated respectively with the local normalization,
the conservation of the global vorticity distribution and of the energy.
Critical points are solutions of: 
\begin{equation}
\forall\ \delta\rho\quad\delta\mathcal{S}-\lambda\delta\mathcal{E}-\int_{\Sigma}\mathrm{d}\sigma\ \alpha\delta D_{\sigma}-\int_{\mathcal{D}}\mathrm{d}\mathbf{r}\ \zeta\delta N=0\ ,\label{eq:critical_points}
\end{equation}
where first variations are taken with respect to $\rho$. This leads
to $\rho=N\exp\left({\lambda\sigma\psi\left(\mathbf{r}\right)-\alpha(\sigma)}\right)$
where N is determined by the normalization constraint $\left(\int\mathrm{d}\sigma\ \rho=1\right)$.
A first result is that equilibrium macrostates characterized by a
functional relation between potential vorticity and streamfunction:
\begin{equation}
\bar{q}=\frac{\int_{\Sigma}\mathrm{d}\sigma\ \sigma e^{\lambda\sigma\psi\left(\mathbf{r}\right)-\alpha(\sigma)}}{\int_{\Sigma}\mathrm{d}\sigma\, e^{\sigma\lambda\psi\left(\mathbf{r}\right)-\alpha\left(\sigma\right)}}=g\left({\psi}\right).\label{eq:q-psi_equilibre}
\end{equation}
We thus conclude that equilibrium macrostates are steady solutions of the quasi-geostrophic dynamics. This is a very clear explanation of the reason why most freely evolving turbulent quasi-geostrophic flows are attracted by steady states.

It can be shown that $g$ is a monotonic and bounded function of ${\psi}$
for any global distribution $\gamma(\sigma)$ and energy $E$. These
critical points can either be entropy minima, saddle or maxima. To
find statistical equilibria, one needs then to select the entropy
maxima. 

When the domain has symmetries, additional
invariants must be taken into account. We will discuss in the next section the important example of angular momentum for flows on a sphere. 

At this point, two different approaches could be followed. The first
one would be to consider a given small scale distribution $\gamma(\sigma)$
and energy $E$, and then to compute the statistical equilibria corresponding
to these parameters. In practice, especially in the geophysical context,
one does not have empirically access to the microscopic vorticity
distribution, but rather to the $q-\psi$ relation (\ref{eq:q-psi_equilibre})
of the large scale flow. The second approach, followed in the remaining
of this chapter, is to study statistical equilibria corresponding
to a given class of $q-\psi$ relations. The precise mathematical relation between these two approaches is discussed in  \citep{bouchet2008}.

\section{Zonal structures as low energy states\label{sec:low_energy_states}}

The main interest of the MRS statistical mechanics approach is to build phase diagrams for the large flow structure. This allows to describes phase transition  associated with drastic changes in the flow topology when key parameters are changed. Depending on the problem at hand, the parameter can be the energy, the circulation, the linear or the angular momentum of the flow, or any other conserved quantity. The energy plays a particular role. Indeed,  when the flow admits symmetries, there may exist for sufficiently high energies  bifurcation associated with the breaking of this symmetry. We will present in the following an illustration of this effect for the breaking of the zonal symmetry. 
%For instance, we will present example of bifurcations in these phase diagrams obtained by varying the energy, the circulation, the linear or the angular momentum of the flow. 

We consider in the following the quasi-geostrophic dynamics  on a beta plane ($\eta_{d}=\beta y$). For the sake of simplicity, we assume a barotropic model ($R=+\infty$), but all the results can be generalized to arbitrary Rossby radius of deformation $R$. General results on MRS equilibria are usually extremely difficult to obtain, and complete phase diagrams can be described only case by case, depending for instance on the domain geometry, and on a particular $q-\psi$ relation. The first phase diagrams of MRS equilibria computed analytically were presented by Chavanis and Sommeria in the case of the Euler equations in a closed domain \citep{chavanissommeria1996}.    Many other results have been obtained in the last two decades, and a lot remains to be done in that direction.   We will focus in the remaining of this section on  MRS states associated with a linear $ q - \psi $ relation for various geometries. The main conclusion is that whenever the domain geometry is invariant by translation, low energy states on a beta plane are purely zonal, and this zonal symmetry can be broken at higher energy.  {The case of a tanh-like $q-\psi$ relation  will be discussed in section \ref{sec:Rsmall}}.

\subsection{The case of a closed domain on a beta plane}

We consider first the case of a closed domain, with $\psi=0$ at the boundary.  MRS equilibria characterized by a linear $\overline{q}-\psi$ relation are solutions of the following variational problem:
\begin{equation}
Z(E,\Gamma)=\max_{\overline{q}}\left\{-\mathcal{Z}[\overline{q}] \ | \ \mathcal{E}[\overline{q}]=E,\ \mathcal{G}[\overline{q}]=\Gamma\right\}, \label{eq:var_closed}
\end{equation}
which means that we look for the state $\overline{q}$ that minimize the macroscopic potential enstrophy $\mathcal{Z}=(1/2) \int\mathrm{d}\mathbf{r} \ \overline{q}^2 $ while satisfying the energy constraint  $\mathcal{E}[\overline{q}]=E$ and the circulation constraint  $\mathcal{G} [\overline{q}]= \int\mathrm{d}\mathbf{r}\overline{q}$.  The solutions of the variational problem (\ref{eq:var_closed}) are actually statistical equilibria of the MRS theory, as proved in \cite{bouchet2008} ; more precisely a subclass of all possible statistical equilibria. We limit our discussion to this subclass for simplicity (we note that perturbative corrections of the class of equilibrium macrostates with linear potential vorticity-stream function relations have been studied in several recent papers \cite{loxleynadiga2013,bouchetcorvellec2010}). Here we only note that critical points of  Eq. (\ref{eq:var_closed}) satisfies $\delta \mathcal{Z} +\mathcal{\lambda} \mathcal{E} +\gamma\mathcal{G}=0$, which  leads to $\overline{q}=\lambda \psi +\gamma$. We have introduced  Lagrange multipliers $\lambda(E,\Gamma)$ and $\gamma(E,\Gamma)$ associated respectively with  the energy and the circulation constraints. The variational problem can be solved analytically, because it involves the maximization of a quadratic functional (the entropy) in the presence of a quadratic constraint (the energy) and a linear constraint (the circulation).  A complete solution of this variational problem for any closed domain is given in \cite{venaillebouchet2011a}. Here we focus on a rectangular domain, and discuss the main properties of the phase diagram.  %sufficiently stretched in the $x$-direction. The corresponding phase diagram is shown Fig. \ref{Fofonoff}. 

We note that the variational problem  Eq. (\ref{eq:var_closed}) involves only two parameters $E>0$ and $\Gamma$, which can be related to the two coefficients of the linear  $\overline{q}-\psi$ relation. One can always rescale time unit such that $E=1$. Then the potential circulation is transformed into  $\Gamma/E^{1/2}$. This shows that only one parameter is needed to describe the phase diagram. To simplify the discussion we assume $\Gamma/E^{1/2}\ne 0$ in the following.

\begin{figure}
%\figurebox{\columnwidth}{0.38\columnwidth}{phase_diag_xfig.eps} 
\figurebox{\columnwidth}{}{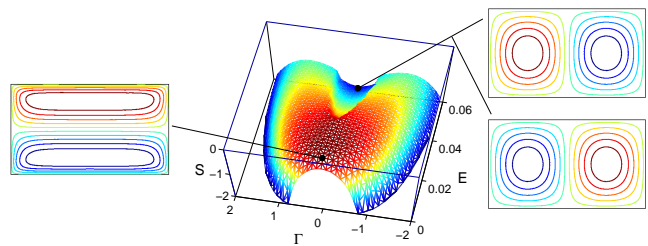} 
\caption{{Phase diagrams of MRS statistical equilibrium states
 characterized by a linear $q-\psi$ relationship for an elongated rectangle.  $S(E,\Gamma)$ is the equilibrium entropy, $E$ is the
energy and $\Gamma$ the circulation. Low energy states are the 
Fofonoff solutions. High energy states have a very different structure (a dipole), see \cite{venaillebouchet2011a} for more details). \label{fig:Fofonoff} }}
\end{figure}

When the parameter $\Gamma/E^{1/2}$ is sufficiently large,  there is a unique solution to the variational problem Eq.  (\ref{eq:var_closed}). This corresponds to a low energy limit, in which case the flow structure is given by the celebrated Fofonoff solution  \citep{fofonoff1954}, obtained by assuming  $\lambda \gg L^{-2}$, where $L$ is
the domain size. In this limit, the Laplacian term in (\ref{dir})
is negligible in the domain bulk. Then $\psi \approx \left(\beta y-\gamma \right)/\lambda$,
which corresponds to a weak westward flow, as illustrated figure \ref{fig:Fofonoff}.
Strong recirculating eastward jets occur at northern and southern
boundaries, where the Laplacian term is no more negligible. We conclude that low energy states are zonal in the domain bulk. 

Let us first assume that the domain aspect ratio is sufficiently stretched in the $x$ direction, just as in Fig. \ref{fig:Fofonoff}. 
When the parameter $\Gamma/E^{1/2}$ is decreased below  a critical value, there is a second order phase transition that breaks the symmetry $x\leftrightarrow -x$, and the variational problem admits two solutions at each point $E,\Gamma$ of the phase diagram. For sufficiently low values of  $\Gamma/E^{1/2}$  the flow has a dipolar structure  just as in figure \ref{fig:Fofonoff}. Remarkably, these high energy states are no more zonal in the domain bulk. 

The phase diagram is a bit different for other domain aspect ratio: when the domain is sufficiently stretched in the $y$-direction, there is a second order phase transition that breaks the $y\leftrightarrow -y$  symmetry, and high energy states have a dipolar structure in the $y$ direction. When the domain aspect ratio is close to one, there is no second order phase transition and high energy states have a monopole structure. In any case, high energy states are not zonal in the domain bulk.  Moreover, these high energy states correspond to the MRS equilibrium states of the Euler equation computed in  \citep{chavanissommeria1996}.

The original low energy  Fofonoff solution was obtained independently from statistical
mechanics considerations. The linear $q-\psi$ relationship was chosen
as a starting point to compute analytically the flow structure. Because
both the Salmon--Holloway--Hendershott statistical theory \citep{salmonhollowayetal1976}
and the Bretherton--Haidvoguel minimum enstrophy principle \citep{brethertonhaidvogel1976}
did predict a linear relationship between vorticity and streamfunction,
it has been argued that statistical equilibrium theory predicts the
emergence of the classical Fofonoff flows, which had effectively been
reported in numerical simulations of freely decaying barotropic flows
on a beta plane for some range of parameters \citep{wangvallis1994}.
It is shown in \citep{bouchet2008} that all those theories
are particular cases of the MRS statistical mechanics theory. It has
then been actually proven in \cite{venaillebouchet2009,venaillebouchet2011a} that the classical Fofonoff solutions are
indeed MRS statistical equilibria in the limit of low energies, but that a richer variety of flow structure exist.

\subsection{The case of a channel on a beta plane}

We explained in the previous section that MRS equilibria at low energy are zonal in the domain bulk. However, the presence of a western boundary breaks the zonal symmetry of the solution.  This motivates the study of MRS equilibria in a zonal channel on a beta plane  with periodic boundary condition in the $x$-direction. For the sake of simplicity, we choose there to impose the value of the streamfunction $\psi=0$ at the northern and the southern boundary. In that case the dynamics is fully described by Eq.  (\ref{QG}-\ref{dir}), and admits the same conservation laws as in the previous subsection. The equilibrium states are therefore solutions of the variational problem Eq. (\ref{eq:var_closed}).  Then the phase diagrams described in the previous subsection also apply for the channel case. Only the structure of the flow associated with each equilibrium state is a bit different, because of the periodicity of the domain in the $x$-direction.  

To simplify the discussion we assume $\Gamma/E^{1/2}\ne 0$. For sufficiently large values of $\Gamma/E^{1/2}$, the flow is purely zonal, with a weak westward flow in the interior and recirculating eastward jets at the northern and southern boundary.  This corresponds to the Fofonoff solution in an open channel. When the parameter $\Gamma/E^{1/2}$ is decreased,  two cases appear: 
\begin{itemize}
\item either the domain is sufficiently
stretched in the $y$ direction so that its aspect ratio $L_{x}/L_{y}$
is smaller than a critical value $\tau_c$. In that case the flow structure
remains purely zonal, whatever the energy 
\item  or  the aspect ratio  $L_{x}/L_{y}$ is larger than the critical value $\tau_c$, in which case there exists a second
order phase transition above which a dipolar state is observed, just
as in figure \ref{fig:Fofonoff}.
\end{itemize}
This example shows the existence of bifurcation at high energy that breaks the zonal symmetry of the MRS equilibria. 

For the sake of simplicity, we made the arbitrary choice  $\psi=0$ at the domain boundaries in the discussion above. This choice breaks the translational symmetry of the problem. 

In the general case, the channel geometry  is a much more complicated problem. Indeed, the value of the stream-function at the southern boundary can always be set to zero, but the value of the stream function at the northern boundary is a constant that must be specified by an additional equation, see e.g. \cite{pedlosky1982}. This equation is given by the conservation of the circulation along one of the boundaries. In addition, the system becomes in that case invariant by translation, and there is another invariant associated with this symmetry, namely the linear momentum  $\mathcal{L}\equiv\int_{\mathcal{D}}\mathrm{d}\mathbf{r}\overline{q}$. 

Consequently, the  variational problem to solve in order to find MRS equilibria associated with a linear $q-\psi$ relation is  
\begin{equation}
S(E,\Gamma)=\max_{\overline{q}}\left\{-\mathcal{Z}[\overline{q}] \ | \ \mathcal{E}[\overline{q}]=E,\ \mathcal{G}^{\pm}[\overline{q}]=\Gamma^{\pm} ,\ \mathcal{L}[\overline{q}]=L\right\}\label{eq:var_channel}
\end{equation}
where $\mathcal{G}^{\pm}$ is the circulation along each boundary.  
Since $\mathcal{L}$ and $\mathcal{G}^{\pm}$  are linear functionals, the computation of the
equilibrium state associated with a linear $q-\psi$ relation is analogous
to the computation of the equilibria in a closed domain presented in the previous section.
However,  there is a richer phase diagrams in the channel case 
due the presence of  additional  parameters. There exists to our knowledge no complete computation of the phase diagrams in the literature for this problem, except in \cite{corvellec2012} who compute equilibrium states in the presence of a topography that breaks the translational invariance (in which case $\mathcal{L}$ is not a dynamical invariant). Indeed, the computation and the physical understanding of phase diagrams for MRS equilibria is still an active subject, and many progresses toward a better physical understanding of this problem in various geometries or for other flow models are expected in the coming years.

\subsection{Zonal flows on a sphere as equilibrium macrostates.}

Computation of statistical equilibria on a sphere have first been
addressed by \citep{frederiksensawford1980} in the context of energy-enstrophy
theories, see also \citep{majdawang2006,limnebus2007} and references
therein. Recently, complete phase diagrams of the equilibrium states
have been explicitly computed by \citep{herbertdubrulleetal2012a} in
the case of linear $\overline{q}-\psi$ relations, taking into account the conservation
of the vertical projection of the angular momentum  
\begin{equation}
\mathcal{L}_z\equiv\int_{\mathcal{D}}\mathrm{d}\mathbf{r}\overline{q} \cos\theta,\label{eq:linear_momentum_sphere}
\end{equation}
where $\theta$ is the latitude.  More precisely, they solved the variational problem

\begin{equation}
S(E,\Gamma)=\max_{\overline{q}}\left\{-\mathcal{Z}[\overline{q}] \ | \ \mathcal{E}[\overline{q}]=E,\ \mathcal{L}_z[\overline{q}]=L_z\right\} \ .\label{eq:var_sphere}
\end{equation}

Note that the circulation on a sphere necessarily vanishes. This is why it does not appears as a parameter in the variational problem. Similarly to the case of the variational problem (\ref{eq:var_closed}), as discussed below formula (\ref{eq:var_closed}), the solution to (\ref{eq:var_sphere}) is a simple subclass of MRS equilibrium macrostates. Note also that  time unit can be rescaled such that $E =1$. This means that the linear momentum is transformed into  $L_z/E^{1/2}$. This rescaling shows that just as in the close domain case of the first subsection, the phase diagrams can be described using only one parameter.

\begin{figure}
%\figurebox{\columnwidth}{0.51\columnwidth}{sphere_herbert.eps} 
\figurebox{\columnwidth}{}{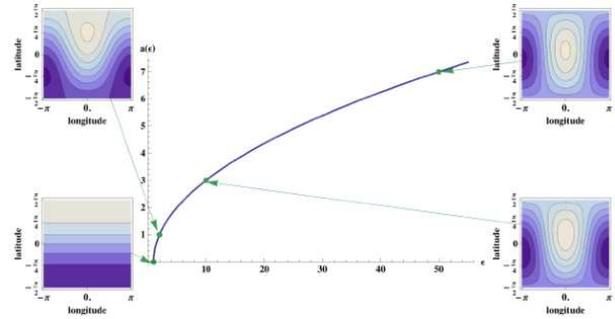} 
\caption{{ Equilibrium states on a sphere for linear $q-\psi$ relations. The $x$-axis represents the energy of the
flow divided by the minimum admissible energy for a given value of
the vertical projection of the angular momentum. See \citep{herbertdubrulleetal2012a} 
for more details. \label{fig_sphere} }}
\end{figure}

The main conclusion of \cite{herbertdubrulleetal2012a} concerning the
structure of the equilibrium states is summarized on  Fig. \ref{fig_sphere}. No solution exist when $L_z/E^{1/2}$ is larger than a critical value $A_c$. The flow structure is purely zonal for  $L_z/E^{1/2}=A_c$, and the zonal symmetry is broken whenever $L_z/E^{1/2}<A_c$ , in which case the solution includes a propagating dipole that  dominates the flow structure  for sufficiently small  $L_z/E^{1/2}$.
 
 The role of additional conservation laws given by the projection of the momentum in other directions has been addressed by \citep{herbert2013additional},
who found richer phase diagrams including equilibrium states with a quadripolar structure. 

\section{Zonal jets in the limit of small Rossby radius of deformation\label{sec:Rsmall}}

Another interesting and physically relevant limit that allows for analytical description of equilibrium macrostates is the limit of
Rossby deformation radius $R$ much smaller than the size of the domain
($R\ll L$). This limit is also relevant to discuss potential vorticity homogenization (section \ref{sec:homogenization}). It also provides a further example of a transition from zonal to non-zonal structures at large energy (section \ref{sec:Jupiter} and figure \ref{Fig:phase_top}). It has also been used to explain the appearance of strong localized jets (section \ref{sec:strong_jets}), to model the Great Red Spot and other Jovian vortices (section \ref{sec:Jupiter}), and to discuss the structure of strong eastward jets such as the Gulf-Stream and the Kuroshio in the oceans (section \ref{sec:Gulf Stream and Kuroshio}).

In the case $R\ll L$,  the nonlinearity of the potential vorticity-stream function
relation becomes essential and we show below that the variational problem of the statistical
theory is analogous to the Van-Der-Waals Cahn Hilliard model that
describes phase separation and phase coexistence in usual thermodynamics.

In the context of quasi-geostrophic flow, this leads to interfaces separating
phases characterized by a different value of coarse-grained  potential vorticity, and corresponds
to sub-domains in which the potential vorticity is homogenized. The
interfaces correspond to strong localized jets of typical width $R$.
This limit is relevant to describe equilibrium state with strong jet structures.

We will discuss application of these results to the description of
the large scale organization of oceanic currents in inertial region,
dominated by turbulence, such as the eastward jets like the Gulf Stream
or the Kuroshio extension (the analogue of the Gulf Stream in the
Pacific ocean). In that case the length $L$ could be thought as the
ocean basin scale $L\simeq1000\, km$ (see section \ref{sec:Gulf Stream and Kuroshio}),
and $R$ is the internal Rossby deformation radius with $R\simeq50\, km$
at mid-latitude. We will also discuss how these results can be relevant
for describing some of Jupiter's features, like for instance the Great
Red Spot of Jupiter, which is a giant anticyclone of typical size
$L\simeq20,000\, km$, whith a Rossby radius $R\simeq1000\, km$.
This last case is particularly interesting since it allows to show
that when varying an external parameter, an equilibrium state can
switch from a zonal jet state to a non zonal, coherent vortex state.

\subsection{The Van der Waals--Cahn Hilliard model of first order phase transitions
\label{sub:Van Der Waals}}

The Van der Waals--Cahn Hilliard model is a classical model of thermodynamics
and statistical physics that describes the coexistence of phase in
usual thermodynamics. We give in the following subsections a heuristic
description of this model based on physical arguments, but more details
can be found inst \citep{modica1987} and \citep{bouchet2001}.
Application to the quasi-geostrophic case will be discussed in the
next subsection. The Van der Waals--Cahn Hilliard model involves the
minimization of a free energy 
\begin{equation}
\mathcal{F}=\int_{\mathcal{D}}\mathrm{d}{\bf \mathbf{r}}\,\left[\frac{R^{2}\left(\nabla\phi\right)^{2}}{2}+f(\phi)\right],
\end{equation}
with a linear constraint $\mathcal{A}\left[\phi\right]=\int_{\mathcal{D}}\mathrm{d}{\bf r}\,\phi$:
\begin{equation}
F=\min\left\{ \mathcal{F}\left[\phi\right]\,\,\left|\,\,\mathcal{A}\left[\phi\right]=-B\right.\right\}, \label{eq:Van Der Waals Cahn Hilliard}
\end{equation}
where $\phi$ is the non-dimensional order parameter (for instance
the non-dimensionality local density), and $f\left(\phi\right)$ is
the non-dimensional free energy per unit volume. We consider the limit
$R\ll L$ where $L$ is a typical size of the domain. We assume that
the specific free energy $f$ has a double well shape (see figure
\ref{fig_f}), characteristic of a phase coexistence related to a
first order phase transition. For a simpler discussion, we also assume
$f$ to be even; this does not affect the properties of the solutions
discussed bellow.

\subsubsection{First order phase transition and phase separation \label{sub:phase separation} }

\begin{figure}
\figurebox{0.7\columnwidth}{}{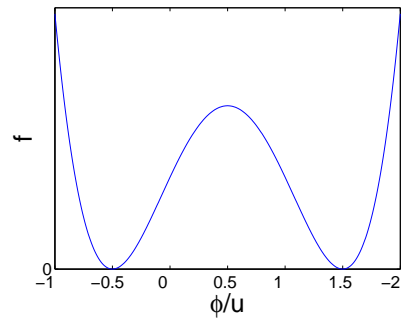} 
\caption{{ The double well shape of the specific free energy $f\left(\phi\right)$.  \label{fig_f}}} 
\end{figure}

At equilibrium, in the limit of small $R$, the function $f\left(\phi\right)$
plays the dominant role. In order to minimize the free energy, the
system will tend to reach one of its two minima (see figure \ref{fig_f}).
These two minima correspond to the value of the order parameters for
the two coexisting phases, the two phases have thus the same free
energy.

\begin{figure}
\figurebox{0.7\columnwidth}{}{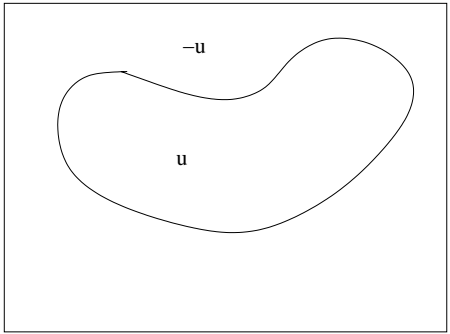} 
\caption{{ At zeroth order, $\phi$ takes the two values $\pm u$
on two sub-domains $A_{\pm}$. These sub-domains are separated by
strong jets.  \label{fig_domaine}}}
\end{figure}

 Without the constraint $\mathcal{A}=-B$ in Eq. (\ref{eq:Van Der Waals Cahn Hilliard}), the two uniform solutions $\phi=u$
or $\phi=-u$ would clearly minimize $\mathcal{F}$: the system would
have only one phase. Because of the constraint $\mathcal{A}$, the
system has to split into sub-domains: part of it with phase $\phi=u$
and part of it with phase $\phi=-u$. In a two dimensional space,
the area occupied by each of the phases are denoted $A_{+}$ and $A_{-}$
respectively. They are fixed by the constraint $\mathcal{A}$ by the
relations $uA_{+}-uA_{-}=-B$ and by $A_{+}+A_{-}=1$ (where $1$
is the total area). A sketch of a situation with two sub-domains each
occupied by one of the two phases is provided in figure \ref{fig_domaine}.

Up to now, we have neglected the term $R^{2}\left(\nabla\phi\right)^{2}$
in the functional (\ref{eq:Van Der Waals Cahn Hilliard}). In classical
thermodynamics, this term is related to non-local contributions to
the free energy. These non-local interactions become
negligible for scales larger than $R$.  Their contribution is therefore limited to the interface between two different phases.  
 
We know from observations of the physical phenomena of coarsening and 
phase separations, that the system has a tendency to form
larger and larger sub-domains. We thus assume that such sub-domains
are delimited by interfaces, with typical radius of curvature $r$
much larger than $R$.   As described in next
sections, this explains the interface structure and interface shape.

\subsubsection{The interface structure}

At the interface, the value of $\phi$ changes rapidly, on a scale
of order $R$, with $R\ll r$. What happens in the direction along
the interface can thus be neglected at leading order. To minimize
the free energy (\ref{eq:Van Der Waals Cahn Hilliard}), the interface
structure $\phi(\zeta)$ needs thus to minimize a one dimensional
variational problem along the normal to the interface coordinate $\zeta$:
\begin{equation}
F_{int}=\min\left\{ \int\mathrm{d}\zeta\,\left[\frac{R^{2}}{2}\left(\frac{d\phi}{d\zeta}\right)^{2}+f(\phi)\right]\right\} .\label{eq:Variational Free Energy Unit lenght}
\end{equation}
where $F_{int}$ a the free energy per unit length of the interface.

We see that the two terms in (\ref{eq:Variational Free Energy Unit lenght})
are of the same order only if the interface has a typical width of
order $R$. We rescale the length by $R$: $\zeta=R\tau$. The Euler-Lagrange
equation of (\ref{eq:Variational Free Energy Unit lenght}) gives ${d^{2}\phi}/{d\tau^{2}}={df}/{d\phi}$.
%\begin{equation}
%\frac{d^{2}\phi}{d\tau^{2}}=\frac{df}{d\phi}.\label{jet}
%\end{equation}
Making an analogy with mechanics, if $\phi$ would be a particle position,
$\tau$ would be the time, this equation would describe the
conservative motion of the particle in a potential $V=-f$. From the
shape of $f$ (see figure \ref{fig_f}) we see that the potential
has two bumps (two unstable fixed points) and decays to $-\infty$
for large distances. In order to connect the two different phases
in the bulk, on each side of the interface, we are looking for solutions
with boundary conditions $\phi\rightarrow\pm u$ for $\tau\rightarrow\pm\infty$.
It exists a unique trajectory with such limit conditions: in the particle
analogy, it is the trajectory connecting the two unstable fixed points
(homoclinic orbit).

This analysis shows that the interface width scales like $R$. Moreover,
after rescaling the length, one clearly sees that the free energy
per length unit (\ref{eq:Variational Free Energy Unit lenght}) is
proportional to $R$: $F_{int}=eR,$ where $e>0$ could be computed
as a function of $f$ \citep{bouchetsommeria2002,venaillebouchet2011b}.

{Note that the emergence of sharp jets separating regions of homogenised potential vorticity can also be predicted using cascade phenomenology\cite{venaillenadeauetal2014} , and  the spontaneous emergence of such flow patterns  has been reported by \cite{arbicflierl2003,venaillenadeauetal2014} in the framework of two-layer quasi-geostrophic flow with very strong bottom friction.}

\subsubsection{The interface shape: an isoperimetrical problem \label{sub:The-interface-shape}}

In order to determine the interface shape, we come back to the free
energy variational problem (\ref{eq:Van Der Waals Cahn Hilliard}).
In the previous section, we have determined the transverse structure
of the interface, by maximizing the one dimensional variational problem
(\ref{eq:Variational Free Energy Unit lenght}). The total free energy
to minimize is thus 
\begin{equation}
\mathcal{F}=LF_{int}=eRL,\label{eq:Free energy length}
\end{equation}
where we have implicitly neglected contributions of relative order
$R/r$.

In order to minimize the free energy (\ref{eq:Free energy length}),
we thus have to minimize the length $L$. We must also take into account
that the areas occupied by the two phases, $A_{+}$ and $A_{-}$ are
fixed, as discussed in section \ref{sub:phase separation}. We thus
look for the curve with the minimal length, that bounds a surface
with area $A_{+}$
\begin{equation}
\min\left\{ eRL\left|\mbox{Area}=A_{+}\right.\right\} .\label{eq:Isoperimetric}
\end{equation}

The solution of the problem (\ref{eq:Isoperimetric})  leads to $\frac{eR}{r}=\alpha$
where $\alpha$ is a Lagrange parameter associated with the conservation
of the area. This proves that $r$ is constant along the interface:
solutions are either circles or straight lines. The law $\frac{eR}{r}=\alpha$
is the equivalent of the Laplace law in classical thermodynamics,
relating the radius of curvature of the interface to the difference
of pressure inside and outside of the bubble. In the following sections,
we see how this applies to the description of statistical equilibria
for quasi-geostrophic flows, describing vortices and jets.%

\subsection{Quasi-geostrophic statistical equilibria and first order phase transitions
\label{sub:QG strong jet generql}}

The first discussion of the analogy between statistical equilibria
in the limit $R\ll L$ and phase coexistence in usual thermodynamics,
in relation with the Van-Der-Waals Cahn Hilliard model is given in
\citep{bouchet2001,bouchetsommeria2002}. This analogy has
been put on a more precise mathematical ground by proving that the
variational problems of the MRS statistical mechanics and the Van-Der-Waals
Cahn Hilliard variational problem are indeed related \citep{bouchet2008}.
More precisely, any solution to the variational problem: 
\begin{equation}
F=\min\left\{ \mathcal{F}\left[\phi\right]\,\,\left|\,\,\mathcal{A}\left[\phi\right]=-B\right.\right\} \label{eq:Variational Van-Der-Waals Topography}
\end{equation}
with $\mathcal{A}\left[\phi\right]=\int_{\mathcal{D}}\mathrm{d}{\bf \mathbf{r}}\,\phi$ and 
\begin{equation}
\mathcal{F}=\int_{\mathcal{D}}\mathrm{d}{\bf \mathbf{r}}\,\left[\frac{R^{2}\left(\nabla\phi\right)^{2}}{2}+f\left(\phi\right)-\eta_{d}\phi\right]
\end{equation}
 is a MRS equilibria of the quasi-geostrophic
equations (\ref{QG}), provided that where $\phi$ is interpreted as a rescaled streamfunction $\psi=R^2\phi$.

Critical points of (\ref{eq:Variational Van-Der-Waals Topography})
are  solutions of $\delta\mathcal{F}-\alpha\delta\mathcal{A}=0$,
for all $\delta\phi$, where $\alpha$ is the Lagrange multiplier
associated with the constraint $\mathcal{A}$. Using a part integration and the relation $q=R^{2}\Delta\phi-\phi+Rh$
yields  $\delta\mathcal{F}=\int\mathbf{\mathrm{d}r}\ \left(f^{\prime}(\phi)-\phi-q\right)\delta\phi$ and $\delta\mathcal{A}=\int\mathrm{d}\mathbf{r}\ \delta\phi$.  The critical points
satisfy  $q=f^{\prime}\left(\frac{\psi}{R^{2}}\right)-\frac{\psi}{R^{2}}-\alpha$.
 We conclude that this equation is the same as (\ref{eq:q-psi_equilibre}),
provided that $f^{\prime}\left(\frac{\psi}{R^{2}}\right)=g(\beta\psi)+\frac{\psi}{R^{2}}-\alpha$.

In the case of an initial distribution $\gamma$ with only two values
of the potential vorticity: $\gamma(\sigma)=\left|\mathcal{D}\right|\left(a\delta(\sigma_{1})+(1-a)\delta(\sigma_{2})\right)$,
only two Lagrange multipliers $\alpha_{1}$ and $\alpha_{2}$ are
needed, associated with $\sigma_{1}$ and $\sigma_{2}$ respectively,
in order to compute $g$, equation (\ref{eq:q-psi_equilibre}). In
that case, the function $g$ is exactly $\tanh$ function. There exists
in practice a much larger class of initial conditions for which the
function $g$ would be an increasing function with a single inflexion
point, similar to a $\tanh$ function, especially when one considers
the limit of small Rossby radius of deformation. \citep{bouchetsommeria2002,venaillebouchet2011b}
give physical arguments to explain why it is the case for Jupiter's
troposphere or oceanic rings and jets.

When $g$ is a $\tanh$-like function, the specific free energy $f$
has a double well shape, provided that the inverse temperature $\beta_{t}$
is negative, with sufficiently large values.

\paragraph{Topography and anisotropy}

The topography term $\eta_{d}$ in (\ref{eq:Variational Van-Der-Waals Topography})
is the main difference between the Van-Der-Waals Cahn Hilliard functional
(\ref{eq:Van Der Waals Cahn Hilliard}) and the quasi-geostrophic
variational problem (\ref{eq:Variational Van-Der-Waals Topography}).
We recall that this term is due to the beta plane approximation and
a prescribed motion in a lower layer of fluid (see section \ref{sub:quasi-geostrophic Model}).
This topographic term provides an anisotropy in the free energy. Its
effect will be the subject of most of the theoretical discussion in
the following sections. For that purpose we assume that this term
scale with the Rossby radius of deformation $R$: $\eta_{d}=R\tilde{\eta}_d$.
With this scaling, the topography term will not change the overall
structure at leading order: there will still be phase separations
in sub-domains, separated by an interface of typical width $R$, as
discussed in section \ref{sub:Van Der Waals}. We now discuss the
dynamical meaning of this overall structure for the quasi-geostrophic
model.

\paragraph{Potential vorticity homogenization and phase separation}
\label{sec:homogenization}

In the case of the quasi-geostrophic equations, the order parameter
$\phi$ is proportional to the stream function $\psi$: $\psi=R^{2}\phi$.
At equilibrium, there is a functional relation between the stream
function $\psi$ and the macroscopic potential vorticity $q$, given
by Eq. (\ref{eq:q-psi_equilibre}). Then the sub-domains of constant
$\phi$ are domains where the (macroscopic) potential vorticity $q$
is also constant. It means that the level of mixing of the different
microscopic potential vorticity levels are constant in those sub-domains.
We thus conclude that the macroscopic potential vorticity is homogenized
in sub-domains that corresponds to different phases (with different
values of potential vorticity), the equilibrium being controlled by
an equality for the associated mixing free energy.

\paragraph{Strong jets and interfaces}
\label{sec:strong_jets}

In section \ref{sub:The-interface-shape}, we have described the interface
structure. The order parameter $\phi$ varies on a scale of order
$R$ mostly in the normal to the interface direction, reaching constant
values far from the interface. Recalling that $\phi$ is proportional
to $\psi$, and that $\mathbf{v} = \mathbf{e}_{z} \times \nabla \psi$,
we conclude that: 
\begin{enumerate}
\item The velocity field is nearly zero far from the interface (at distances
much larger than the Rossby deformation radius $R$). Non zero velocities
are limited to the interface areas. 
\item The velocity is mainly directed along the interface. 
\end{enumerate}
These two properties characterize strong jets. In the limit $R\ll L$,
the velocity field is thus mainly composed of strong jets of width
$R$, whose path is determined from an isoperimetrical variational
problem.

\subsection{Are the Gulf-Stream and the Kuroshio currents close to statistical
equilibria? \label{sec:Gulf Stream and Kuroshio}}

\begin{figure} 
\figurebox{\columnwidth}{}{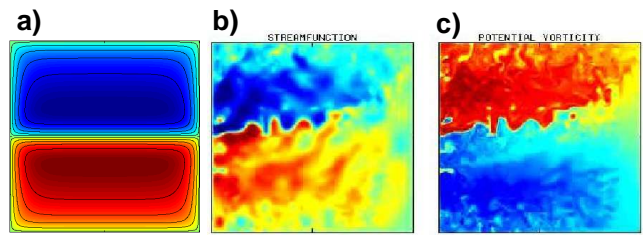} 
\caption{ { b) and c) represent respectively a snapshot of the
streamfunction and potential vorticity (red: positive values; blue:
negative values) in the upper layer of a three layers quasi-geostrophic
model in a closed domain. Both figures are taken from numerical
simulations by P. Berloff. a) Streamfunction predicted by statistical
mechanics. \label{Fig:frontBerloff} }}
\end{figure}

We have mentioned that statistical equilibria, starting from the Van-Der-Waals
Cahn Hilliard functional (\ref{eq:Variational Van-Der-Waals Topography}),
may model physical situations where strong jets, with a width of order
$R$, bound domains of nearly constant potential vorticity. This is
actually the case of the Gulf Stream in the North Atlantic ocean or
of the Kuroshio extension in the North Pacific ocean. This can be
inferred from observations, or this is observed in high resolution
numerical simulations of idealized wind driven mid-latitude ocean,
see for instance figure \ref{Fig:frontBerloff} (see \citep{berloffhoggetal2007}
for more details).

We address the following problem: is it possible to find a class of
statistical equilibria solution of the with a strong mid-basin eastward
jet similar to the Gulf Stream of the Kuroshio, in a closed domain?
We analyze therefore the class of statistical equilibria of the 1-1/2
layer model which are minima of the Van-Der-Waals Cahn Hilliard variational
problem (\ref{eq:Variational Van-Der-Waals Topography}). We ask whether
it exists solutions to in a bounded domain (let say a rectangular
basin) with strong mid-basin eastward jets. At the domain boundary,
we fix $\phi=0$ (which using $\phi=R^{2}\psi$ turns out to be an
impermeability condition). We further assume that the equivalent topographic
term can be writen on the form 
\begin{equation}
\eta_{d}=R\tilde{\beta}y,\label{eq:effective_top}
\end{equation}
which includes the beta effect and the effect of a deep zonal flow
($\eta_{d}=\beta y+\psi_{d}/R^{2}$ with $\psi_{d}=-U_{d}y$). As
discussed in chapter \ref{sub:Van Der Waals}, with these hypothesis,
there is phase separation in two subdomains with two different levels
of potential vorticity mixing, provided that the function $f\left(\phi\right)$
has a double well shape. These domains are bounded by interfaces (jets)
of width $R$. In view of the applications to mid-basin ocean jets,
we assume that the area $A_{+}$ occupied by the value $\phi=u$ is
half of the total area of the domain (this amounts to fix the total
potential vorticity $\Gamma_{1}$). The question is to determine the
position and shape of this interface.

\subsubsection{Eastward jets are statistical equilibria of the quasi-geostrophic
model without topography\label{sub:Eastward-jets-without_topography}}

\begin{figure}
\figurebox{\columnwidth}{}{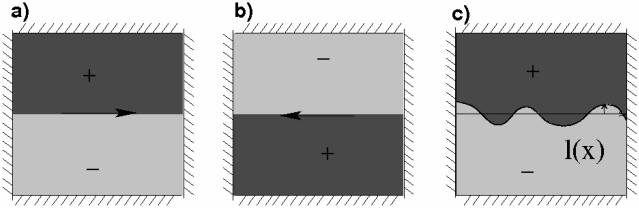} 
\caption{{ a) Eastward jet: the interface is zonal, with positive
potential vorticity $q=u$ on the northern part of the domain. b)
Westward jet: the interface is zonal, with negative potential vorticity
$q=-u$ in the northern part of the domain. c) Perturbation of the
interface for the eastward jet configuration.  \label{fig_pvfronts}}} 
\end{figure}

The value $\phi=\pm u$ for the two coexisting phases is not compatible
with the boundary condition $\phi=0$. As a consequence, there exists
a boundary jet (or boundary layer) in order to match a uniform phase
$\phi=\pm u$ to the boundary conditions. Just like inner jets, treated
in section \ref{sec:low_energy_states}, these jets contribute to
the first order free energy, which gives the jet position and shape.
We now treat the effect of boundary layer for the case $h=0$ ($\widetilde{\beta}=0$
in this case). As explained in section \ref{sub:The-interface-shape},
the jet free energy is the only contribution to the total free energy.

We first quantify the unit length free energy, $F_{b}$, for the boundary
jets. Following the reasoning of section \ref{sub:The-interface-shape},
$F_{b}$ is given my the minimum of   $\int d\zeta\,\left[\frac{R^{2}}{2}\frac{d^{2}\phi}{d\zeta^{2}}+f(\phi)\right]$. 
 This expression is the same as (\ref{eq:Variational Free Energy Unit lenght}),
the only difference is the different boundary conditions: it was $\phi\rightarrow_{\zeta\rightarrow+\infty}u$
and $\phi\rightarrow_{\zeta\rightarrow-\infty}-u$, it is now $\phi\rightarrow_{\zeta\rightarrow+\infty}u$
and $\phi\left(0\right)=0$. Because $f$ is even, one easily see
that a boundary jet is nothing else than half of a interior domain
jet. Then $F_{b}=\frac{e}{2}R$,
where  $e$ is defined in subsection \ref{sub:The-interface-shape}.
By symmetry, a boundary jet matching the value $\phi=-u$ to $\phi=0$
gives the same contribution.
 Finally, the first order free energy is given by  $\mathcal{F}=eR\left(L+\frac{L_{b}}{2}\right)$,
where $L_{b}$ is the boundary length. Because the boundary length
$L_{b}$ is a fixed quantity, the free energy minimisation amounts
to the minimisation of the interior jet length. The interior jet position
and shape is thus given by the minimisation of the interior jet length
with fixed area $A_{+}.$ We recall that the solutions to this variational
problem are interior jets which are either straight lines or circles
(see section \ref{sub:The-interface-shape}).

In order to simplify the discussion, we consider the case of a rectangular
domain of aspect ratio $\tau=L_{x}/L_{y}$. Generalisation to an arbitrary
closed domain could also be discussed. We recall that the two phases
occupy the same area $A_{+}=A_{-}=\frac{1}{2}L_{x}L_{y}$. We consider
three possible interface configurations with straight or circular
jets: 
\begin{enumerate}
\item the zonal jet configuration (jet along the $x$ axis) with $L=L_{x}$, 
\item the meridional jet configuration (jet along the $y$ axis with $L=L_{y}$, 
\item and an interior circular vortex, with $L=2\sqrt{\pi A_{+}}=\sqrt{2\pi L_{x}L_{y}}$. 
\end{enumerate}
The minimisation of $L$ for these three configurations shows that
the zonal jet is a global minimum if and only if $\tau<1$. The criterion
for the zonal jet to be a global MRS equilibrium state is then $L_{x}<L_{y}$.
We have thus found zonal jet as statistical equilibria in the case
$h=0$.

An essential point is that both the Kuroshio and the Gulf Stream are
flowing eastward (from west to east). From the relation $\mathbf{v}=\mathbf{e}_{z}\times\nabla\psi$,
we see that the jet flows eastward ($v_{x}>0$) when $\partial_{y}\psi<0$.
Recalling that $\phi=R^{2}\psi$, the previous condition means that
the negative phase $\phi=-u$ has to be on the northern part of the
domain, and the phase $\phi=u$ on the southern part. From (\ref{dir}),
we see that this corresponds to a phase with positive potential vorticity
$q=u$ on the northern sub-domains and negative potential vorticity
$q=-u$ on the southern sub-domain, as illustrated in the panel (a)
of figure (\ref{fig_pvfronts}).

Looking at the variational problems (\ref{eq:Variational Van-Der-Waals Topography}),
it is clear that in the case $\eta_d=0$ and $f$ even, the minimisation
of $\phi$ is invariant over the symmetry $\phi\rightarrow-\phi$.
Then solutions with eastward or westward jets are completely equivalent.
Actually there are two equivalent solutions for each of the case 1,
2 and 3 above. However, taking into account the beta effect 
will break this symmetry. 

\subsubsection{Addition of a topography \label{sub:Eastward-jets-with_topography}}

We now consider the effect of a small effective beta effect as in
Eq. (\ref{eq:effective_top}) . This choice of a topography amplitude
scaling with $R$ is considered in order to treat the case where the
contribution of the effective beta effect appears at the same order
as the jet length contribution.  Following the arguments of the previous subsections,
we minimize 
\begin{equation}
\mathcal{F}=RH_{0}+R\left(eL-2u\int_{A_{+}}\mathrm{d}{\bf \mathbf{r}}\,\tilde{\beta}y\right),\label{eq:Energy_Libre_Jet_Effet_Beta}
\end{equation}
with a fixed area $A_{+}$.
The jet position is a critical point of this functional: $e/r-2u\tilde{\beta}y_{jet}=\alpha$, where $\alpha$ is the Lagrange
parameter associated with the constraint  on $A_+$ and $y_{jet}$ the latitude of the jet. We conclude that
zonal jets (curves with constant $y_{jet}$ and $r=+\infty$) are
solutions to this equation for $\alpha=-2uR\tilde{\beta}y_{jet}$.
Eastward and westward jets described in the previous section are still
critical points of entropy maximisation.

\paragraph{With a negative effective beta effect, eastward jets are statistical
equilibria.}

We first consider the case $\widetilde{\beta}<0$. This occurs when
the zonal flow in the lower layer is eastward and sufficiently strong
($U_{d}>R^{2}\beta$). If we compute the first order free energy (\ref{eq:Energy_Libre_Jet_Effet_Beta})
for both the eastward and the westward mid-latitude jet, it is easy
to see that in order to minimise $\mathcal{F}$, the domain $A_{+}$
has to be located at the lower latitudes: taking $y=0$ at the interface,
the term $-2u\int_{A_{+}}d^{2}{\bf r}\,\tilde{\beta}y=u\tilde{\beta}L_{x}L_{y}/4$
gives a negative contribution when the phase with $\phi=u$ (and $q=-u$)
is on the southern part of the domain ($A_{+}=(0,L_{x})\times(-\frac{L_{y}}{2},0)$).
This term would give the opposite contribution if the phase $\phi=u$
would occupy the northern part of the domain. Thus the statistical
equilibria is the one with negative streamfunction $\phi$ (corresponding
to positive potential vorticity $q$) on the northern part of the
domain. As discussed in the end of section \ref{sub:Eastward-jets-without_topography}
and illustrated on figure \ref{fig_pvfronts}, panel (b), this is
the case of an eastward jet.

Thus, we conclude that taking into account an effective negative beta-effect
term at first order breaks the westward-eastward jet symmetry. When
$\widetilde{\beta}<0$, statistical equilibria are flows with mid-basin
eastward jets.

\paragraph{With a positive effective beta effect, westward jets are statistical
equilibria.}

Let us now assume that the effective beta coefficient is positive.
This is the case when $U_{d}<R^{2}\beta$, i.e. when the lower layer
is either flowing westward, or eastward with a sufficiently low velocity.
The argument of the previous paragraph can then be used to show that
the statistical equilibrium is the solution presenting a westward
jet.

\paragraph{With a sufficiently small effective beta coefficient, eastward jets
are local statistical equilibria.}

We have just proved that mid-basin eastward jets are not global equilibria
in the case of positive effective beta effect. They are however critical
points of entropy maximisation. They still could be local entropy
maxima. We now consider this question: are mid-basin strong eastward
jets local equilibria for a positive effective beta coefficient? In
order to answer, we perturb the interface between the two phases,
while keeping constant the area they occupy, and compute the free
energy perturbation.

The unperturbed interface equation is $y=0$, the perturbed one is
$y=l(x)$, see figure \ref{fig_pvfronts}. Qualitatively, the contributions
to the free energy $\mathcal{F}$ (\ref{eq:Energy_Libre_Jet_Effet_Beta}),
of the jet on one hand and of the topography on the other hand, are
competing with each other. Any perturbation increases the jet length
$L=\int dx\ \sqrt{1+\left(\frac{dl}{dx}\right)^{2}}$ and then increases
the second term in equation (\ref{eq:Energy_Libre_Jet_Effet_Beta})
by $\delta\mathcal{F}_{1}=Re\int dx\,\left(dl/dx\right)^{2}$. Any
perturbation decreases the third term in equation (\ref{eq:Energy_Libre_Jet_Effet_Beta})
by $\delta\mathcal{F}_{2}=-2Ru\tilde{\beta}\int dx\,\ l^{2}$.

We suppose that $l=l_{k}\sin\frac{k\pi}{L_{x}}x$ where $k\ge1$ is
an integer. Then $\delta\mathcal{F}=-2u\tilde{\beta}+e\left(\frac{k\pi}{L_{x}}\right)^{2}$. Because we minimize $\mathcal{F}$, we want to know if any perturbation
leads to positive variations of the free energy. The most unfavorable
case is for the smallest value of $k^{2}$, i.e. $k^{2}=1.$ Then
we conclude that eastward jets are metastable states (local entropy maxima) when $\widetilde{\beta}<\widetilde{\beta}_{cr}=\frac{1}{2}\frac{e}{u}\frac{\pi^{2}}{L_{x}^{2}}$.

\paragraph{The critical zonal extension for metastable mid basin eastward jets. }

The previous result can also be interpreted in terms of the domain
geometry, for a fixed value of $\widetilde{\beta}$. Eastward jets
are local entropy maxima if $ L_{x}<L_{x,cr}=\pi\sqrt{\frac{e}{2u\widetilde{\beta}_{cr}}}$.
Let us evaluate an order of magnitude for $L_{x,cr}$ for the ocean
case, first assuming there is no deep flow ($U_{d}=0$). Then $R\tilde{\beta}$
is the real coefficient of the beta plane approximation. Remembering
that a typical velocity of the jet is $U\sim uR$, and using $e\sim u^{2}$ gives $L_{x,cr}\approx\pi\sqrt{\frac{U}{\beta_{cr}}}$, \citep{venaillebouchet2011b}.
This length is proportional to the Rhine's' scale of geophysical fluid
dynamics \citep{vallis2006}. For jets like the Gulf Stream, typical
jet velocity is $1\ m.s^{-1}$ and $\beta\sim10^{-11}\ m^{-1}.s^{-1}$
at mid-latitude. Then $L_{cr}\sim300\ km$. This length is much smaller
than the typical zonal extension of the inertial part of the Kuroshio
or Gulf Stream currents. We thus conclude that in a model with a quiescent
lower layer and the beta plane approximation, currents like the Gulf
Stream or the Kuroshio are not statistical equilibria, and they are
not neither close to local statistical equilibria.

Taking the oceanic parameters ($\beta=\ 10^{-11}\ m^{-1}s^{-1}$,
$R\sim50\ km$), we can estimate the critical eastward velocity in
the lower layer $U_{d,cr}=5\ cm\ s^{-1}$ above which the strong eastward
jet in the upper layer is a statistical equilibria. It is difficult
to make further conclusions about real mid-latitude jets; we conjecture
that their are marginally stable. This hypothesis of marginal stability
is in agreement with the observed instabilities of the Gulf-Stream
and Kuroshio current, but overall stability of the global structure
of the flow.

\subsubsection{Is equilibrium statistical mechanics relevant to describe mid-basin eastward jets ? }

The early work of Fofonoff and the equilibrium statistical mechanics
of geophysical flows presented in this review are often referred to
as the inertial approach of oceanic circulation, meaning that the
effect of the forcing and the dissipation are neglected.

Ocean dynamics is actually much influenced by the forcing and the
dissipation. For instance the mass flux of a current like the Gulf
Stream is mainly explained by the Sverdrup transport. Indeed in the
bulk of the ocean, a balance between wind stress forcing and beta
effect (the Sverdrup balance) lead to a meridional global mass flux
(for instance toward the south on the southern part of the Atlantic
ocean. This fluxes is then oriented westward and explain a large part
of the Gulf Stream mass transport. This mechanism is at the base of
the classical theories for ocean dynamics, see e.g. \citep{pedlosky1998}.
Because it is not an conservative process, the inertial approach does
not take this essential aspect into account. Conversely, the traditional
theory explains the Sverdrup transport, the westward intensification
and boundary current, but gives no clear explanation of the structure
of the inertial part of the current: the strongly eastward jets.

Each of the classical ocean theory or of the equilibrium statistical
mechanics point of view give an incomplete picture, and complement
each other. Another interesting approach consider the dynamics from
the point of view of bifurcation theory when the Reynolds number (or
some other controlled parameters) are increased. These three types
of approaches seem complimentary and we hope they may be combined
in the future in a more comprehensive non-equilibrium theory.

\subsection{From zonal jets to circular vortex: application to Jovian vortices. }
\label{sec:Jupiter}

\begin{figure}
\figurebox{\columnwidth}{}{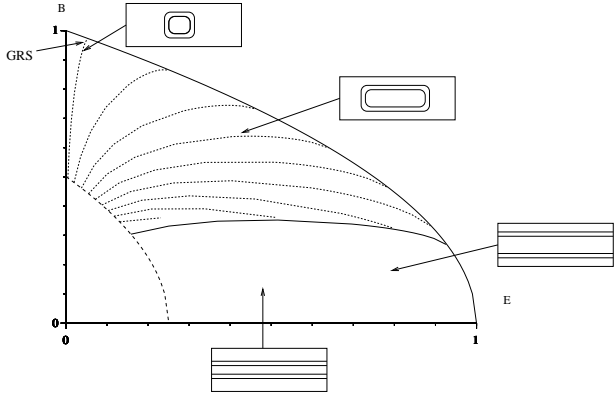} 
\caption{{Phase diagram of the statistical equilibrium states in a channel with quadratic topography.  The parameters $E$ and $B$ are respectively the energy and an asymmetry parameter for the potential vorticity distribution. The inner solid line corresponds to a phase transition,
between vortex and straight jet solutions. The dash line corresponds
to the limit of validity of the small deformation radius hypothesis.
The dot lines are constant vortex aspect ratio lines. See \citep{bouchetdumont2003} for more details. \label{Fig:phase_top} }}
\end{figure}

The previous subsection was focused on the possible existence of zonal
jets as statistical equilibriums states. The same statistical mechanics
ideas can be used to describe various Jovian vortices, see e.g. \citep{bouchetsommeria2002,bouchetvenaille2012}
and ocean rings \citep{venaillebouchet2011b}. A natural question
is to be able to predict wether an initial condition will self-organize
into a vortex or into a zonal structures. Figure \ref{Fig:phase_top}
shows a phase diagram for statistical equilibria with Jupiter like
topography in a channel, see \citep{bouchetdumont2003}
for more details. This illustrates the power of statistical mechanics:
with only few parameters characterizing statistical equilibria, one can  reproduce all
the features of Jupiter's troposphere, from circular white ovals,
to the Great Red Spot and cigar shaped Brown Barges. Here, these parameters are
the energy $E$ and a parameter $B$ defined Eq. (\ref{eq:Variational Van-Der-Waals Topography}), related to the asymmetry between positive
and negative potential vorticity.  As seen on figure \ref{Fig:phase_top}, statistical mechanics
predicts a phase transition form straight jets to vortices when increasing the asymmetry parameter $B$. The reduction
of the complexity of turbulent flow to only a few order parameters
is the main interest and achievement of a statistical mechanics theory.

\section{Discussion and conclusion}

The theoretical study of the self-organization of geostrophic turbulent
flows into zonal jets has been addressed in this chapter  based on statistical mechanics
methods. The theory predicts the output of the long time evolution
of a class of turbulent flows described by the quasi-geostrophic equations.
We have explained in which cases zonal flow may be interpreted as statistical
equilibrium, and how the zonal symmetry may be broken when some key
parameters such as the total energy are varied. On this chapter, emphasize
has been placed on examples with available analytical treatment in
order to favor a better understanding of the physics and dynamics.
We have described phase diagrams for closed domain, channel and spherical
geometry in the case of linear $q-\psi$ relations. {Note that there exist several works generalising those results to weakly non-linear $q-\psi$ relations, see e.g. \cite{majdawang2006,bouchetvenaille2012,qimarston2013} and references therein.} We have then discussed the limit of small
Rossby radius of deformation which allows to describe strong jets
at the interface of homogenized regions of potential vorticity, in which case the $q-\psi$ relation is a tanh-like function. This
has been  applied to make quantitative models ocean jets like the Gulf-Stream. %The power of the statistical mechanics approach is to be able to predict the large scale flow structure as the outcome of turbulent dynamics with only a few key parameters.  

We focused in this chapter on one layer quasi-geostrophic models, and consequently did not discuss statistical mechanics predictions for the vertical structure of geostrophic flows, which may have some importance in the context of the emergence of zonal structures on a beta plane. For instance, statistical mechanics arguments were used in  \cite{venaillevallisetal2012} to interpret numerical simulations of continuously stratified fluids showing that the beta effect favors the tendency towards barotropization of the flow. Other results on statistical mechanics predictions theory for the vertical structure of geostrophic flows are presented in \cite{merryfield1998,venaille2012,herbert2014}.

\subsection{Observed multiple jets are not statistical equilibrium states}

Multiple eastward jets are often observed in the simulation of quasi-geostrophic dynamics, or even in observations in various geophysical contexts (see ?? in this book).  In this chapter, none of the equilibrium state we described presented  this multiple jet structure. 

To our knowledge, all the  multiple jets configurations described in the literature are characterized by a potential vorticity profile increasing with $y$, while the  streamfunction has roughly a sawtooth shape corresponding to a change of sign of the velocity. For instance, in the case of potential vorticity staircases obtained  in (\citep{dritschelmcintyre2008} or \citep{danilovgurarie2004}), the velocity field is eastward at the interface between regions of homogenized potential vorticity, and there is a weaker westward recirculation close in the interior of the homogenized potential vorticity regions. This means that all reported multiple jets are characterized by a non-monotonous $q-\psi$ relation. 

One can show that the $q-\psi$  relation of the critical point of the MRS theory is necessarily a monotonous function. This means that observed multiple jets can not be equilibrium states, nor metastable states of the equilibrium theory. A qualitative interpretation can be given: the zonal jets act as a mixing barrier of potential vorticity in physical space, but also (and consequently) in phase space. The ergodic hypothesis underlying the statistical mechanics equilibrium theory is therefore not satisfied.

\subsection{The role of forcing and dissipation}

Equilibrium
statistical mechanics can be valid only if the effects of forcing
and dissipation can be neglected. This corresponds to two different
kind of situation. 

The first one, as discussed in the original papers
\citep{robert1990,miller1990,robert1991,robertsommeria1991}.
is when the flow is produced by an instability or from a prepared
initial condition, and then evolve over a time scale which is much
smaller than the typical time scales associated to the non-inertial
processes (forcing and dissipation). In geophysical context, this framework is probably the
correct one for the formation of ocean mesoscale eddies,
from the instability of either the Gulf Stream (Gulf Stream rings)
or or the Agulhas current downstream of Cape Agulhas, see e.g. \citep{venaillebouchet2011b}. 
{This is also the situation  relevant to describe freely decaying turbulent flows in numerical simulations, see e.g. \cite{brandsmaassenetal1999,qimarston2013} and references therein. Althought he MRS theory can generally be used  to interpret qualitatively self-organisation phenomena in numerical simulations, two limits can be pointed out. First, the global distribution  of potential vorticity generally changes before relaxation toward equilibrium, see e.g. \cite{brandsmaassenetal1999}, and it  is unclear wether this difficulty may be overcome by increasing numerical resolution. Second, statistical equilibrium states are only a sub-class of stable states of the dynamics. Even in a case without small scale dissipation, the system may therefore be trapped into flow structures that are not predicted by the equilibrium theory.}

Most of geophysical and other natural flows are however in another
regime. Very often they have settled down from a very long time to
a statistically stationary solution, where on average forces balance
dissipation. In this case, one can still compare the typical time
scale of inertial organization (usually turnover times, or typical
times for wave propagation) to the forcing and dissipation time scale
(spin up or spin down time scale). If these two time scales are well
separated, then we still expect equilibrium statistical mechanics
to describe at leading order the flow structure, and its qualitative
properties. Usefulness of equilibrium statistical mechanics in this
second framework, for instance close to a phase transition, is illustrated
in \citep{bouchetsimonnet2008}. We nevertheless note a limitation
of equilibrium statistical mechanics in this second framework. It
does not predict which of the set of possible statistical equilibria
(parameterized by the inertial invariants) is actually selected by
the long term effect of forces and dissipation. This should be determined
at next order by computing the vanishingly small fluxes of conserved
quantities.

Still most of ocean and atmosphere dynamics flows, for instance large
scale organization of the atmosphere or the ocean, do not really fulfill
these separation of time scale hypothesis. Then a truly non-equilibrium
statistical mechanics approach has to be considered. This is the subject
of a number of current approaches, using kinetic theory \citep{nardiniguptaetal2012,bouchetmorita2010},
related approaches such as stochastic structural stability theory
(see \citep{farrellioannou2003,srinivasanyoung2011}
and references therein), or cumulant expansions (see \citep{marston2010,marstonconoveretal2008}
and references therein), or instanton theory. Section 6 of the review
by \citep{bouchetvenaille2012} contains a more complete discussion
of such non-equilibrium approaches; whereas the review by \citep{marston2011,lucarinibenderetal2013}
stresses the interest of statistical mechanics for climate applications.

%\section*{Acknowledgments}
%This work was supported through the ANR program STATFLOW (ANR-06-JCJC-0037-01)
%and through the ANR program STATOCEAN (ANR-09-SYSC-014), and partly
%by DoE grant DE-SC0005189 and NOAA grant NA08OAR4320752. 

\bibliography{equilibrium.bib}
\bibliographystyle{plain}

\end{document}